# A Multi-Technique Study of $C_2H_4$ Adsorption on a Model Single-Atom $Rh_1$ Catalyst

Chunlei Wang [1]*, Panukorn Sombut[1], Lena Puntscher[1], Manuel Ulreich[1], Jiri Pavelec[1], David Rath[1], Jan Balajka[1], Matthias Meier[1,2], Michael Schmid[1], Ulrike Diebold[1], Cesare Franchini[2,3], and Gareth S. Parkinson[1]

[1]Institute of Applied Physics, TU Wien, Vienna, Austria

[2]Faculty of Physics, Center for Computational Materials Science, University of Vienna, Vienna, Austria

[3]Dipartimento di Fisica e Astronomia, Università di Bologna, Bologna, Italy

**ABSTRACT**

Single-atom catalysts are potentially ideal model systems to investigate structure-function relationships in catalysis, if the active sites can be uniquely determined. In this work, we study the interaction of $C_2H_4$ with a model $Rh/Fe_3O_4(001)$ catalyst that features 2-, 5-, and 6-fold coordinated Rh adatoms, as well as Rh clusters. Using multiple surface-sensitive techniques in combination with calculations of density functional theory (DFT), we follow the thermal evolution of the system and disentangle the behavior of the different species. $C_2H_4$ adsorption is strongest at the 2-fold coordinated $Rh_1$ with a DFT-determined adsorption energy of –2.26 eV. However, desorption occurs at lower temperatures than expected because the Rh migrates into substitutional sites within the support, where the molecule is more weakly bound. Adsorption at the 5-fold coordinated Rh sites is predicated to –1.49 eV, but the superposition of this signal with that from small Rh clusters and additional heterogeneity leads to a broad $C_2H_4$ desorption shoulder in TPD above room temperature.

## 1. Introduction

Supported single-atom catalysts (SACs) have garnered much attention due to their cost efficiency, catalytic activity, and high selectivity for various reactions. Significant efforts have been devoted to understanding the reaction mechanisms in SAC,[1-5] but providing a clear elucidation of the structure-activity relationship is complicated by the inability to determine the local bonding environment of the active atoms. Moreover, the stability of the active centers during reactions is still a subject of debate, and it is difficult to determine whether small clusters form under reaction conditions.[6-10] Traditionally, researchers have utilized model catalysts based on well-defined single crystals under ultrahigh vacuum (UHV) conditions to study structure-property relationships in catalysis[11-14] However, few model systems exist in which single atoms remain stable on model supports at reaction temperatures. To date, most model studies of SAC have been limited to the adsorption and reaction of inorganic molecules such as CO, $O_2$, and water.[15-17] Here, we select a representative olefin, ethylene ($C_2H_4$), and study its interaction with a model Rh/$Fe_3O_4$ catalyst featuring single atoms in various configurations.

Olefins serve as vital reactants in many industrial processes such as polymerization or hydroformylation of aldehyde synthesis. A clear understanding of these catalytic reaction pathways is important for the promotion of catalytic properties. Recently, Guo et al. reported an in-situ visualization of ethylene polymerization by scanning tunneling microscopy (STM) on a carburized iron model catalyst, which provides a direct evidence for this growth process.[18] The hydroformylation reaction is typically performed homogeneously in solution utilizing Wilkinson's catalyst, but recently, oxide-supported Rh single-atom catalysts have been shown to exhibit remarkable catalytic performance.[19-21] An important aspect of this reaction is the co-adsorption of CO and $C_2H_4$, so an atomic-scale understanding of the $C_2H_4$ adsorption behavior on $Rh_1$ would be timely.

In this paper, we utilize a single-crystal $Fe_3O_4$(001) support that can stabilize a variety of different metals, and where the coordination can be tuned by the preparation conditions.[22,23] Surface-sensitive techniques such as STM, temperature programmed desorption (TPD), and X-ray photoelectron spectroscopy (XPS) are employed to explore how the ethylene interacts with Rh in different configurations. We found that ethylene adsorption on 2-fold $Rh_1$ sites leads to the formation of a pseudo-square planar structure, in which the Rh relaxes towards the support forming a weak coordination with subsurface oxygen. When the sample is heated in a TPD experiment, ethylene desorption is accompanied by the evolution of Rh adatoms to a substitutional cation geometry in the support. Desorption from the 5-fold coordinated Rh sites occurs just above room temperature, comparable to desorption from small Rh clusters.

## 2. Experimental and Computational Methods

### Experimental

The experiments were performed on natural $Fe_3O_4$(001) (6×6×1 mm) single crystals purchased from SurfaceNet GmbH. All the samples were cleaned by cycles of sputtering (10 min, 1 keV $Ar^+$ (STM chamber) or Ne$^+$ (TPD/XPS chamber) and annealing (923 K, 20 min). In the final cleaning cycle before measurement, the sample was oxidized by annealing in 2×10$^{-6}$ mbar O$_2$ for 20 min at 923 K. The annealing leads to the growth of new pristine surface layers, and yields the reconstructed ($\sqrt{2}\times\sqrt{2}$)R45° surface.[24] Rh atoms were deposited using an e-beam evaporator (FOCUS), with the flux calibrated using a temperature stabilized quartz microbalance (QCM). One monolayer (ML) is defined as one Rh atom per Fe$_3$O$_4$(001)-($\sqrt{2}\times\sqrt{2}$)R45° surface unit cell, which is equivalent to 1.42×10$^{14}$/cm$^2$.

Two separate UHV systems were utilized during this work: Imaging experiments were performed in a setup that includes a coupled preparation chamber (base pressure $p < 10^{-10}$ mbar) and analysis chamber ($p = 5 \times 10^{-11}$ mbar). STM was performed in the analysis chamber using an Omicron μ-STM at room temperature in constant-current mode using an electrochemically etched W tip. The sample was always positively biased, meaning empty states were imaged. The analysis chamber is also equipped with a nonmonochromatic Al Kα X-ray source and a SPECS Phoibos 100 analyzer for XPS analysis. XPS data acquired here were utilized as a fingerprint to ensure that the sample prepared in the TPD experiments was the same.

The TPD and XPS experiments were conducted in another UHV system optimized to study the surface chemistry of model catalysis.[25] The Fe$_3$O$_4$(001) sample was mounted on a Ta backplate with a thin gold sheet in between to improve the thermal contact. The sample is cooled by a liquid-He flow cryostat (base temperature ~40 K) and is heated by the resistive heating of the Ta backplate. The vacuum system is equipped with a home-built molecular beam source, which delivers reactants with a calibrated flux (equivalent to the impingement rate at 2.66 × 10$^{-8}$ mbar) and a top-hat profile to the sample with a 3.5 mm diameter.[25,26] We give gas doses in Langmuir units, 1 L is defined as 1× 10$^6$ torr s. The Rh was deposited using the same procedure as in the STM experiments described above. C$_2$H$_4$ was used for TPD experiments. A quadrupole mass spectrometer (Hiden HAL 3F PIC) is used in a line-of-sight geometry for TPD experiments, analyzing desorption signal at mass 27, not 28 for C$_2$H$_4$, to avoid any background from trace levels of CO and N$_2$ in the chamber. A monochromatized Al/Ag twin anode X-ray source (Specs XR50 M, FOCUS 500) and a hemispherical analyzer (Specs Phoibos 150) are used for XPS measurements. A complete description of the TPD / XPS chamber design is provided in ref.[25]

**Computational Details**

The Vienna *ab initio* Simulation Package (VASP) was used for all DFT calculations.[27] Using the projector augmented wave method to handle the near-core regions[28,29]. The plane-wave basis set cutoff energy was set to 550 eV. The calculations were performed using the generalized gradient approximation method with the Perdew-Burke-Ernzerhof (PBE) functional to describe electronic exchange and correlation.[30] Dispersion terms are included according to the D3 Becke-Johnson method.[31] An effective on-site Coulomb repulsion term U$_{eff}$ = 3.61 eV was used for the 3*d* electrons of the Fe atoms.[32,33] The convergence crieterion was an electronic energy change of 10$^{-6}$ eV per step, and forces acting on ions smaller than 0.02 eV/Å. Calculations were performed with the experimental magnetite lattice parameter (a = 8.396 Å) using an asymmetric slab with 13 planes (7 planes with octahedral Fe and 6 with tetrahedral Fe; the bottom 9 planes are fixed and only the 4 topmost planes relaxed) and using the Γ-point only for a large (2√2×2√2)R45° supercell. The slabs were separated by a 14 Å vacuum layer. The average adsorption energy of adsorbed $C_2H_4$ molecules on Rh adatom is computed according to the formula

$$E_{ads} = \left(E_{Rh/Fe_3O_4+nC_2H_4} - \left(E_{Rh/Fe_3O_4} + nE_{C_2H_4}\right)\right)/n$$

where $E_{Rh/Fe_3O_4+nC_2H_4}$ is the total energy of the Rh-decorated $Fe_3O_4$(001) surface with adsorbed $C_2H_4$, $E_{Rh/Fe_3O_4}$ is the total energy of the Rh-decorated $Fe_3O_4$(001) surface, the $E_{C_2H_4}$ represents the energy of $C_2H_4$ molecule in the gas phase, and $n$ is the number of $C_2H_4$ molecules.

## 3. Results

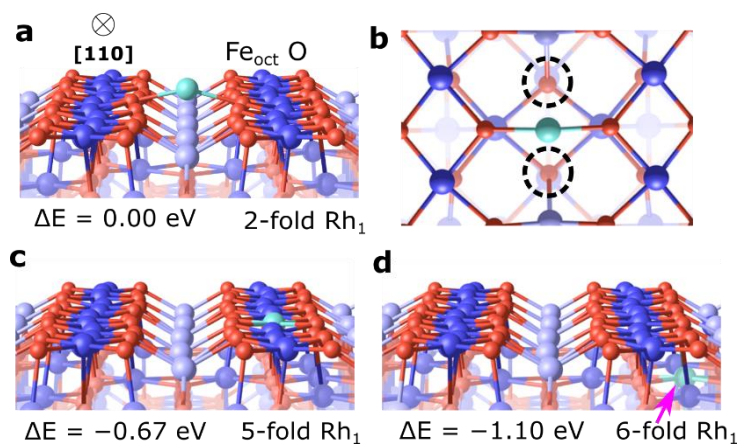

**Figure 1. DFT-determined structure models for the $C_2H_4$/$Rh_1$/$Fe_3O_4$(001) system.** (a) Perspective and (b) top view models of a 2-fold oxygen coordinated $Rh_1$ on $Fe_3O_4$(001) support. The two dashed circles in (b) indicate two equivalent sub-surface oxygen atoms in the support, with which the Rh atoms can form a weak bond. (c) A 5-fold coordinated $Rh_1$ atom in a substitutional cation site and (d) a subsurface Rh site (pink arrow) with 6-fold coordination to lattice oxygen. Oxygen atoms are red in the models, while surface 5-fold coordinated $Fe_{oct}$ atoms are dark blue. Rh is shown as cyan.

The Rh/$Fe_3O_4$(001) model system has been studied recently by different groups.[15,34-36] In general, the reconstructed $Fe_3O_4$(001) surface has been shown to provide four possible configurations for Rh atoms: (1) adatom sites with 2-fold coordination to surface oxygen, (2) surface substitutional sites with 5-fold coordination to oxygen, (3) subsurface substitutional sites with 6-fold coordination to oxygen and (4) Rh clusters, including well-defined dimers.[34] In Figure 1 (a-d), we show the DFT-determined structure and energetics for the isolated $Rh_1$ geometries. The structure shown in Figure 1(a-b) is the 2-fold coordinated $Rh_1$ adatom. It protrudes 0.7 Å above the O atoms in the surface plane, has a Bader charge of 0.7$e$, and the adsorption energy is calculated to be –4.42 eV, referenced to a gas phase Rh atom. This initial geometry is consistent with all other Fe$_3$O$_4$(001)-based model single-atom catalysts studied to date after deposition at room temperature.[22,23,37] The 5-fold Rh$_1$ shown in Figure 1c can be considered as the substitution of a surface iron cation by Rh. This site is 0.67 eV more stable than the 2-fold Rh$_1$, and exhibits a Bader charge of 1.27$e$. The 6-fold coordinated, subsurface Rh$_1$ in a substitutional site, shown in Figure 1d, has an energy 1.10 eV more favorable than the 2-fold Rh$_1$ (–5.52 eV with respect to a single gas-phase Rh) and exhibits a Bader charge of 1.24$e$. Since the 6-fold Rh is subsurface and fully coordinated to O, it cannot bind to a C$_2$H$_4$ molecule. In what follows, we primarily consider adsorption at the 2- and 5-fold Rh sites.

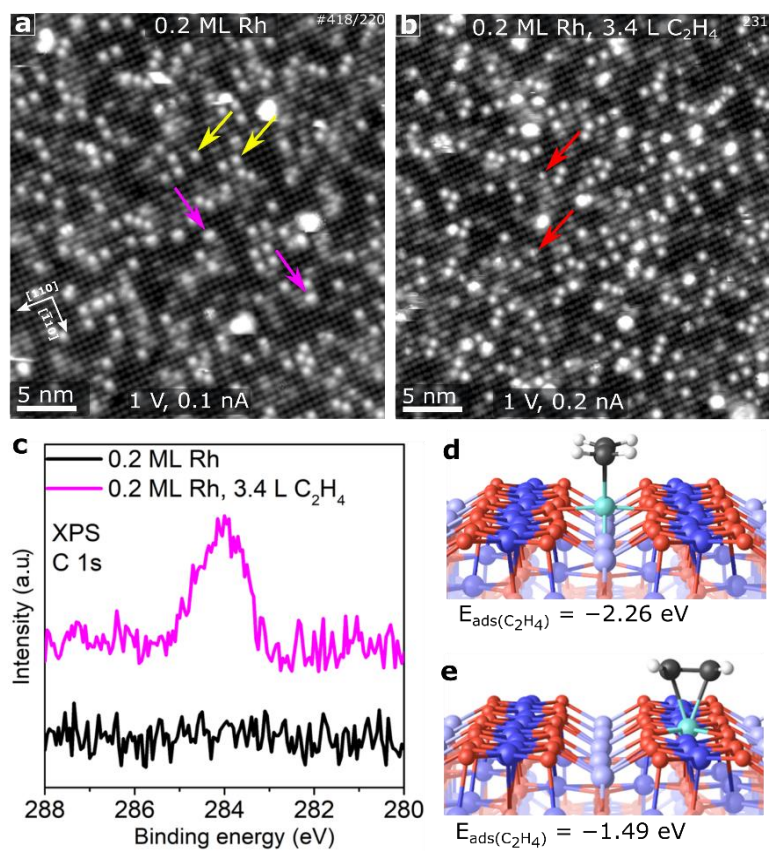

**Figure 2. C$_2$H$_4$ adsorption on Rh$_1$/Fe$_3$O$_4$(001) surface.** STM images of the as-prepared 0.2 ML Rh$_1$/Fe$_3$O$_4$(001) surface (a) before and (b) after 3.4 L C$_2$H$_4$ adsorption. The yellow arrows indicate 2-fold coordinated Rh$_1$ atoms, which are located in-between the iron rows. The pink arrows indicate Rh dimer species, identified in previous work.[34] In image (b) acquired at a different position after adsorption of C$_2$H$_4$, the red arrows indicate protrusions within the surface iron rows, which are due to C$_2$H$_4$ adsorbed on 5-fold coordinated Rh$_1$ atoms. (c) C1s XPS acquired from the 0.2 ML Rh$_1$/Fe$_3$O$_4$(001) sample before (black curve) and after exposure to 3.4 L C$_2$H$_4$ (pink curve), the curves have been shifted vertically for clarity. (d-e) DFT derived minimum energy structures for C$_2$H$_4$ on 2-fold Rh$_1$ and 5-fold Rh$_1$, respectively. The oxygen atoms are red in the models, while surface 5-fold coordinated Fe$_{oct}$ atoms are dark blue. Rh is shown as cyan. The carbon and hydrogen atoms of the ethylene molecule are shown black and white, respectively.

Figure 2 shows an STM image acquired after Rh was deposited directly on the as-prepared $Fe_3O_4(001)$ surface at room temperature via physical vapor deposition (PVD) (see Figure S1 for a corresponding image of the as-prepared surface prior to deposition and details of the typical surface defects observed). The bright rows of protrusions running in the ⟨110⟩ type directions are due to the surface $Fe_{oct}$ atoms, and the bright protrusions (indicated by yellow arrows in STM image of Figure 2) located in-between these rows are assigned to 2-fold coordinated $Rh_1$ adatoms. Note that the surface oxygen atoms are not resolved in the images as they have no density of states in the vicinity of the Fermi level. Nevertheless, their positions are well known from diffractions based experiments.[38] A few small clusters are also observed even at this low coverage, which may be linked to sintering induced by residual $O_2$ following sample preparation.[35] The pink arrows in Figure 2a indicate $Rh_2$ dimers, which are slightly extended along the direction of the surface iron rows, different from single atoms (yellow arrows).[34]

In Figure 2(a-b), we show the 0.2 ML $Rh_1/Fe_3O_4(001)$ surface after exposure to 3.4 L $C_2H_4$ at room temperature. The appearance and apparent height of the protrusions related to the 2-fold Rh are identical to before the exposure within experimental error. Nevertheless, we are confident that these atoms have adsorbed $C_2H_4$ because C 1s XPS data (Figure 2c) obtained from these samples exhibit a peak at ~284 eV consistent with $C_2H_4$ adsorption. Moreover, the area of this peak is approximately double that obtained for a similar coverage of Rh monocarbonyls studied previously[22,34,35]. This suggests that each $Rh_1$ adsorbed a single $C_2H_4$ molecule, which is not visible as a change in STM contrast. In the prior CO experiment, a weak reduction in the apparent height was observed [22,34,35]. Interestingly, the density of adatoms located between the surface Fe rows is approximately 23% higher than before the $C_2H_4$ exposure, suggesting some redispersion of clusters has occurred. We do not observe any species immediately identifiable as the Rh dimer species, so it seems likely that $C_2H_4$ adsorption can break these species apart into 2 adatoms, as was observed previously after CO exposure.[34] In the CO adsorption experiments, we were able to directly observe the CO induced splitting of Rh dimers by in-situ STM imaging. Unfortunately, such an experiment is not possible here because imaging the surface during $C_2H_4$ exposure leads to a dramatic sintering of the system (see Movie S1). This effect does not occur in the absence of the STM tip, which suggests that the tunneling electrons dissociate transiently adsorbed $C_2H_4$ molecules, and the fragments destabilize the Rh atoms. Finally, small bright protrusions directly over the Fe rows (red arrows) are due to adsorption at 5-fold coordinated Rh sites. The coverage of these species is ~12 % of the 2-fold Rh coverage.

The minimum-energy configuration obtained computationally for a $C_2H_4$ molecule adsorbed on a 2-fold $Rh_1$ species on $Fe_3O_4(001)$ is shown in Figure 2d. The carbon-carbon double bond lies parallel to the iron rows, and the adsorption energy is –2.26 eV. An alternative configuration in which the carbon-carbon double bond of $C_2H_4$ lies perpendicular to the iron row (shown in Figure S2) has a weaker adsorption energy of –1.86 eV. The $C_2H_4$ adsorption causes the Rh atom to sink down towards the surface by 0.4 Å, facilitating the formation of a weak bond ($\approx$ 2.36 Å) between the Rh and a subsurface oxygen atom (the relevant O atoms are highlighted by the dashed black circles in Figure 1b). If one considers the Rh-π interaction as a single ligand, the resulting structure creates a pseudo-square planar environment for the Rh atom. This behavior is identical to that observed recently following CO adsorption, so the analogy appears sound.[34,39] As in the CO case, the presence of 2 symmetrically equivalent configurations will allow the system to flip rapidly at room temperature as illustrated in Movie S2 (the DFT determined flipping barrier is ≈ 0.1 eV). The minimum energy configuration of $C_2H_4$ on a 5-fold $Rh_1$ site has the C-C bond perpendicular to the Fe row directions, with an adsorption energy of −1.49 eV (see Figure 2e). This is significantly stronger than for $C_2H_4$ on Fe sites on the pristine surface (−0.54 eV).[40]

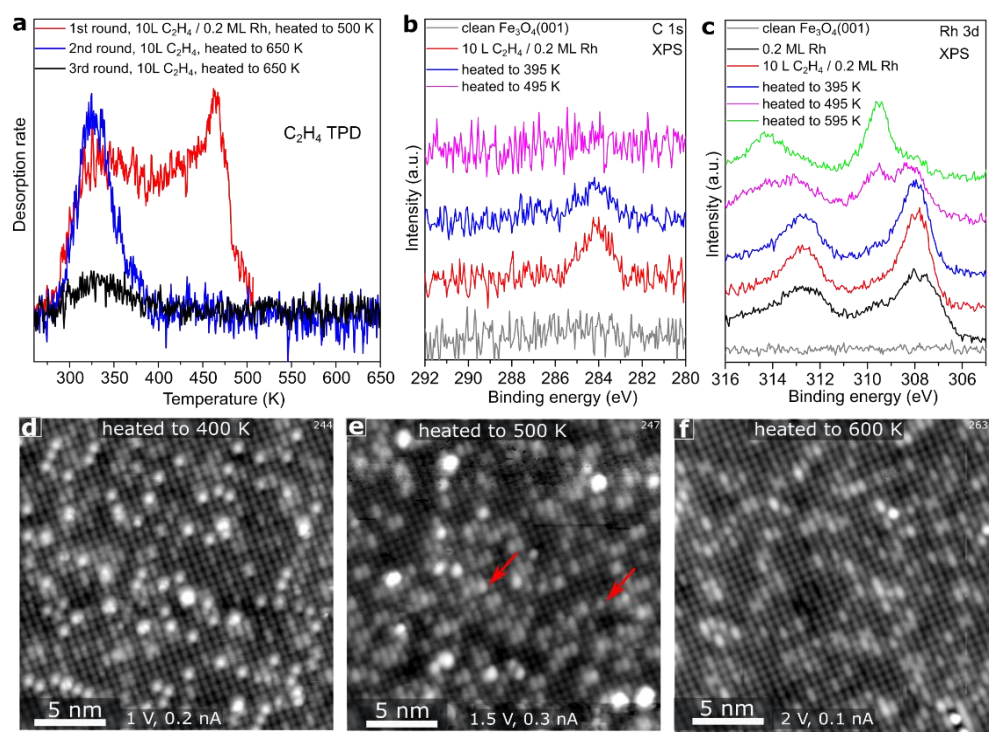

**Figure 3. $C_2H_4$ desorption and Rh evolution.** (a) A series of $C_2H_4$ TPD spectra obtained from 0.2 ML Rh/$Fe_3O_4$(001) sample following exposure to 10 L $C_2H_4$ at 293 K. The TPD was run from 250 K to 500 K in the first round. The sample was then cooled down to 293 K, and a further 10 L $C_2H_4$ adsorbed. In the second and third TPD rounds, the temperature was ramped from 250 K to 650 K. (b-c) XPS of C 1s and Rh 3d collected after different annealing temperatures. The spectrum is collected after sample cooling down room temperature. (d-f) STM images of the sample in Figure 2b, followed by annealing at 400 K, 500 K, and 600 K. The red arrows in (e) indicate 5-fold coordinated $Rh_1$, which may have either formed from 2-fold coordinated $Rh_1$ during annealing or it was present already previously as 5-fold Rh. In any case, it has now lost the adsorbed $C_2H_4$. After annealing at 600 K, panel (f) shows that the sample is similar to a clean $Fe_3O_4$(001) surface as shown in Figure S1.

Further evidence of $C_2H_4$ adsorption at Rh sites comes from TPD and XPS experiments. The red TPD curve in Figure 3 was acquired from a 0.2 ML Rh/$Fe_3O_4$(001) sample after a saturation exposure (see Figure S3) at room temperature. The sample was cooled to 250 K prior to starting the TPD ramp. A continuous $C_2H_4$ desorption spectrum is obtained with a clear peak at 462 K. This suggests that desorption occurs from a variety of different sites with different adsorption energies. To gain a comprehensive understanding of the desorption process, we analyzed the ethylene and Rh using XPS at specific temperatures within the TPD peak. In Figure 3b, the C 1s XPS data reveal that the area of the $C_2H_4$ peak diminishes by ≈ 1/3 when the sample is heated to approximately 395 K. This aligns well with the desorption curve observed in TPD, as less than half of the ethylene is desorbed at that temperature. After annealing to 495 K and 595 K, no carbon species can be detected by C 1s of XPS. We thus conclude that ethylene desorption does not lead to substantial ethylene decomposition and coking of the Rh, and that most, if not all, ethylene desorbed molecularly. This differs from the experience for $C_2H_4$ adsorption and desorption on alumina-supported Pt nanoclusters, where substantial coking is observed.[41]

Turning now to the Rh 3d region (figure 3c) the peak is relatively broad for the as-deposited surface as it contains contributions from small clusters (307 eV), 2-fold Rh adatoms (307.7 eV), 5-fold Rh atoms (308.0 eV), and 6-fold atoms (309.5 eV).[35] The peak becomes sharper after $C_2H_4$ adsorption due to the loss of the low binding energy component. This most likely originates from the redispersion of Rh dimers and small clusters during $C_2H_4$ adsorption that was hypothesized on the basis of the STM images above. No major change in the chemical state and intensity of Rh 3d spectrum is apparent upon heating to 395 K (blue curve), despite desorption of a substantial amount of the $C_2H_4$. The main change observable in STM (Figure 3d) is the disappearance of the molecules that were adsorbed at the 5-fold Rh sites. We will show later that it is likely $C_2H_4$ also desorbed from Rh clusters in this temperature range. Between 395 K and 495 K, there is an evolution of Rh spectrum as the remainder of the ethylene desorbs (Figure 3c, pink curve). The peak splits into two distinct components. The peak at 309.5 eV is attributed to the 6-fold coordinated Rh species, while the peak at 308 eV is due to 5-fold Rh.[35] This assignment is confirmed by the STM image obtained after annealing at 500 K, in which there are essentially no 2-fold Rh adatoms remain (Figure 3e). Instead, many new protrusions are detected (red arrows), these reside within the surface Fe rows and have a similar appearance to the 5-fold coordinated $Rh_1$ shown Figure S4. Some Rh clusters also remain after this annealing step. When the sample is heated further to 600 K, neither 2-fold nor 5-fold Rh remain visible in STM (figure 3f), and the Rh 3d region is dominated by the peak attributed to subsurface 6-fold coordinated Rh atoms (green curve).

In order to further probe the evolution of rhodium species after ethylene desorption, we conducted multiple rounds of $C_2H_4$ TPD experiments as shown in Figure 3a. After the first-round TPD ramp was terminated at 500 K, the sample was cooled to room temperature and re-exposed to $C_2H_4$. Thus, the second TPD curve (shown in blue) is performed on a sample resembling that shown in Figure 3e. In this case, the high temperature peak is missing, and the only $C_2H_4$ desorption is observed to peak at around 330 K. This is further evidence that the most weakly bound $C_2H_4$ in our experiments is located at the 5-fold $Rh_1$ sites and small clusters. A third repeat of the experiment conducted after terminating the previous ramp at 650 K, exhibits a small desorption peak at 330 K, most likely from remaining few clusters or 5-fold Rh that has not diffused to deeper layers yet. The majority of the Rh atoms have got incorporated into the subsurface layers.

Although the desorption profile is rather broad, the peak at 462 K is sufficiently sharp that is possible to perform an analysis of the data to extract the adsorption energy of $C_2H_4$ at the 2-fold adatom sites. For this we utilized a recently developed TPD analysis program described in detail in ref [42], that we had previously used for $C_2H_4$ on the clean $Fe_3O_4$(001) surface.[40] The key assumption in the present work is that the system can be treated as a lattice gas, which essentially means that the molecules do not have any translational degrees of freedom at the desorption temperature. This seems reasonable because the molecules are much more strongly bound at the Rh atom than the surrounding surface, and these sites are clearly still resolved in STM (figure 3d) after heating to 400 K, which is close to the onset of the desorption peak). With this assumption, we compute an experimental adsorption energy of −1.72 eV. The details of TPD data analysis are shown in Figure S5 and in Table S1.

## 4. Discussion

The interaction of $C_2H_4$ with a Rh/$Fe_3O_4$(001) model catalyst was studied using a combination of surface-sensitive techniques and DFT-based calculations. A surprising amount of heterogeneity exists in the initial state of the system considering the apparent homogeneity of the support utilized and the significant energetic differences calculated for the different Rh configurations. The majority of the Rh is initially accommodated in metastable configuration as 2-fold coordinated Rh adatoms. This suggests kinetic stabilization, and thus a barrier to incorporate the Rh into the surface and finally subsurface, which achieves a higher coordination to oxygen. Nevertheless, some Rh gets incorporated already at room temperature,[35] which suggests that there are locations on the surface that facilitate easier incorporation of the Rh. It could be that an incoming Rh atom incident at a particular site within the unit cell encounters a lower barrier for incorporation, or that the proximity to (sub)surface defects eases the process. Unfortunately, it is not possible to tell from the STM images, because Rh incorporation will displace Fe atoms from the subsurface layers where the effect is not visible. However, some insights can be gleaned from a consideration of the thermal evolution of the system. In the absence of $C_2H_4$, annealing the sample to 420 K is sufficient to convert all the 2-fold Rh into more stable configurations, although a mixture of 5- and 6-fold atoms is obtained (Figure S4). $C_2H_4$ adsorption delays the incorporation of the 2-fold Rh into the surface. Together with the fact that a significant proportion of the 5-fold signal clearly remains in XPS after heating to 595 K, we conclude that some sites on the surface stabilize the 5-fold configuration more than others. In addition, one may consider a model where sparse defects such as subsurface Fe vacancie may facilitate the incorporation of Rh as soon as the vacancies get mobile.[43] In such a model, Rh incorporation would depend on whether a Fe vacancy happens to pass under its adsorption site. Thus, the heterogeneity could be partly due to a low concentration of Fe vacancies. Apart from $Rh_1$ species with different coordination, we also observe clusters in the initial state, even at low Rh coverage. Their formation could be the result of random chance, i.e. deposited atoms land in the same place during PVD, but it may also be the result of sintering by residual $O_2$ that remains in the preparation chamber after sample preparation.[35]

The heterogeneity of the model system is clearly seen in the broad TPD data obtained from the as-prepared surface. Nevertheless, we are able to ascertain that the most strongly bound $C_2H_4$ resides at the 2-fold Rh atoms. The adsorption energy obtained from our TPD analysis (−1.72 eV, see Figure S5) is much lower than the DFT computed value of −2.26 eV; even when taking into account that the inclusion of dispersion (D3) over binds small molecules by 0.2-0.3 eV.[22,40] This indicates that the apparent adsorption energy determined from TPD includes the tendency of Rh to assume a higher coordination to oxygen. This suggests that the Rh sheds the $C_2H_4$ in the transition to the 5-fold or 6-fold position. Actually, we saw previously that CO-$Ir_1$ could incorporate as a single entity.[39] Therefore, desorption energy obtained from TPD analysis is not simply the energy difference between 2-fold coordinated Rh with and without adsorbed $C_2H_4$, but it also includes a contribution from the energy gained upon incorporation of Rh during annealing. The link between adsorption and Rh site is also evidenced by the fact that the $C_2H_4$ keeps the Rh in 2-fold sites until the 462 K desorption peak is reached, while Rh would otherwise get incorporated into the surface already at 420 K (Figure S4). The adsorption geometry of $C_2H_4$ on 2-fold Rh is very similar to that obtained previously for CO, where the Rh forms a weak bond to a subsurface oxygen resulting in a pseudo-square planar environment for the Rh.[34] The stability of this environment is in line with the experience from coordination chemistry for Rh(I) systems.

Based on the STM images obtained after heating to 500 K and the appearance of the 2$^{nd}$ TPD round, we conclude that the TPD signal occurring in the 300-400 K temperature regime is due to the $C_2H_4$ desorbing from 5-fold Rh sites. This fits with the much lower adsorption energy calculated by DFT (−1.49 eV). Note we did not specifically analyze the low temperature region because desorption begins immediately at 300 K during the TPD ramp (i.e. the adsorption temperature), which suggests the appearance of the peak in the second TPD round is partly a consequence of the way the experiment was performed. Nevertheless, we conclude that the weakly bound $C_2H_4$ is primarily at the 5-fold Rh sites, and the broadness of the TPD is linked to variations in the environment of the 5-fold Rh (e.g. adjacent defects) as well as the presence of Rh clusters with various sizes. No coking of the system was observed that could be linked to $C_2H_4$ decomposition at the Rh clusters.

The similarity of the adsorption behavior for $C_2H_4$ and CO extends beyond the pseudo-square-planar structure formed at the 2-fold Rh site. In the CO study, we observed that gem dicarbonyl species could not form directly at a 2-fold Rh site, despite the fact that such a configuration should be thermodynamically stable.[34] In the present case, a $Rh_1$ di-ethylene structure is also computed to be possible as shown in Figure S6, and would have a similar structure to the gem dicarbonyl. The differential adsorption energy calculated for the second $C_2H_4$ is only −1.02 eV. Considering the overbinding of the DFT calculations, this might be not enough for stable adsorption of second $C_2H_4$ at room temperature. Nevertheless, we also consider it likely that diethylene formation is prevented by the steric hindrance of the mono $C_2H_4$ species. Interestingly, while Rh gem dicarbonyls could be formed through the decomposition of Rh dimers via a metastable $Rh_2(CO)_3$ configuration,[34] we do not observe any species attributable to Rh di-ethylene in the present study, so it seems that adsorption of 2 ethylene molecules might be sufficient to split the Rh dimers. One possible difference between the CO and the $C_2H_4$ case could be that adsorption of 2 ethylene molecules might be sufficient to split the Rh dimers, and a $Rh_2(C_2H_4)_3$ intermediate (leading to a gem-diethylene) will never be formed. An alternative explanation is that adsorption of three ethylene molecules on a dimer lead to its dissociation, in analogy to the CO case,[34] but at room temperature the resulting $Rh(C_2H_4)_2$ loses its second ethylene soon, due to the insufficient differential adsorption energy. As found for CO, formation of a di-ethylene at the 5-fold Rh site is impossible.

The difference of the adsorption energies of CO and $C_2H_4$ has important consequences for performing hydroformylation of alkenes using SACs, as both reactants are supposed to adsorb at the same Rh site. Clearly, CO will poison the catalyst unless the reaction is performed at a temperature where CO will desorb. The formation of gem-dicarbonyls has been observed by diffuse reflectance infrared Fourier transform spectroscopy in most studies,[21,44] and the reaction is generally performed with a significant excess of $C_2H_4$. The lack of dicarbonyls and di-ethylene forming in our studies likely means that coadsorption of CO and $C_2H_4$ will also not occur in UHV experiments, which suggests that ambient pressure studies will be required to shed more light on this important reaction.

## 5. Conclusions

The adsorption of ethylene on a $Rh/Fe_3O_4(001)$ model catalyst was investigated using surface-sensitive techniques and DFT calculations. The adsorption of ethylene induces a downward relaxation of the 2-fold coordinated $Rh_1$ and leads to a weak coordination with the subsurface oxygen atoms of $Fe_3O_4(001)$. This $C_2H_4$ adsorption (DFT-determined adsorption energy –2.26 eV) results in the formation of a pseudo-square planar configuration for the Rh atom, as found previously for CO. Adsorption at 5-fold coordinated sites is significantly weaker (–1.49 eV). The TPD spectrum of $C_2H_4$ is broad due to the existence of different Rh species, and because $C_2H_4$ desorption from 2-fold $Rh_1$ sites occurs in conjunction with the incorporation of the Rh atom into the surface. All ethylene desorption occurs by 500 K, and Rh tends to incorporate in the subsurface layers of the support where it becomes unavailable for further adsorption.

**Author information**

*Corresponding email: wangc@iap.tuwien.ac.at

**Notes**

The authors declare no competing financial interests.

**Acknowledgements**

LH, GSP, JP, PS, ARA and MM acknowledge funding from the European Research Council (ERC) under the European Union's Horizon 2020 research and innovation programme (grant agreement No. [864628], Consolidator Research Grant 'E-SAC'). This research was funded in part by the Austrian Science Fund (FWF) 10.55776/F81 and 10.55776/Y847. The Vienna Scientific Cluster was used to obtain the computational results. For the purpose of open access, the author has applied a CC BY public copyright licence to any Author Accepted Manuscript version arising from this submission.

**Supporting information**

The supporting information includes STM images of the clean $Fe_3O_4(001)$ surface. STM images and XPS data of $Rh/Fe_3O_4(001)$ before and after annealing at 420 K. Model structures of one and two $C_2H_4$ adsorption on $Rh_1/Fe_3O_4(001)$. A series of $C_2H_4$ TPD results obtained after various $C_2H_4$ exposures at room temperature on a 0.2 ML $Rh/Fe_3O_4(001)$. The details of analysis of the TPD spectra using the TPD program.

Supplementary Information:

# A Multi-Technique Study of C$_2$H$_4$ Adsorption on a Model Single-Atom Rh$_1$ Catalyst

Chunlei Wang [1]*, Panukorn Sombut[1], Lena Puntscher[1], Manuel Ulreich[1], Jiri Pavelec[1], David Rath[1], Jan Balajka[1], Matthias Meier[1,2], Michael Schmid[1], Ulrike Diebold[1], Cesare Franchini[2,3], and Gareth S. Parkinson[1]

[1]Institute of Applied Physics, TU Wien, Vienna, Austria
[2]Faculty of Physics, Center for Computational Materials Science, University of Vienna, Vienna, Austria
[3]Dipartimento di Fisica e Astronomia, Università di Bologna, Bologna, Italy

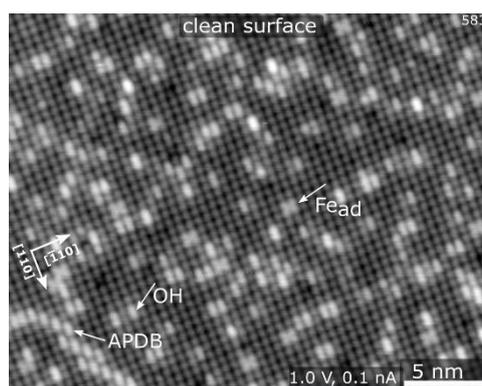

**Figure S1**. The as-prepared Fe$_3$O$_4$(001) clean surface. Scanning tunneling microscopy (STM) image of the reconstructed Fe$_3$O$_4$(001) surface prepared by cycles of sputtering and annealing in a partial pressure of $2 \times 10^{-6}$ mbar O$_2$. The protrusions forming rows in [110] direction are due to pairs of surface Fe atoms. Bright protrusions are related to various surface defects including a surface hydroxyl group (OH), and anti-phase domain boundary in the surface reconstruction (APDB) and an Fe adatom Fe$_{ad}$.[1] Each is labelled within the image.

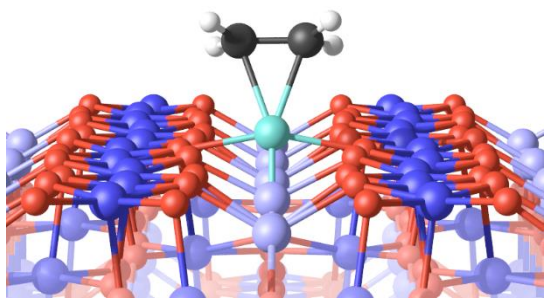

**Figure S2.** An alternative variant of $C_2H_4$ adsorption on the 2-fold coordinated $Rh_1$ determined by DFT calculation. In this variant, the C=C double bond of $C_2H_4$ molecule is perpendicular to the iron rows of $Fe_3O_4$(001) support. The adsorption energy of $C_2H_4$ towards this structure is – 1.86 eV, which is less favorable compared to the $C_2H_4$ adsorption structure in Figure 2d, which features a C=C bond parallel to the surface Fe rows. The oxygen atoms are red in the models, while surface 5-fold coordinated $Fe_{oct}$ atoms are dark blue. Rh is shown as cyan. The carbon and hydrogen atoms of the ethylene molecule are shown black and white, respectively.

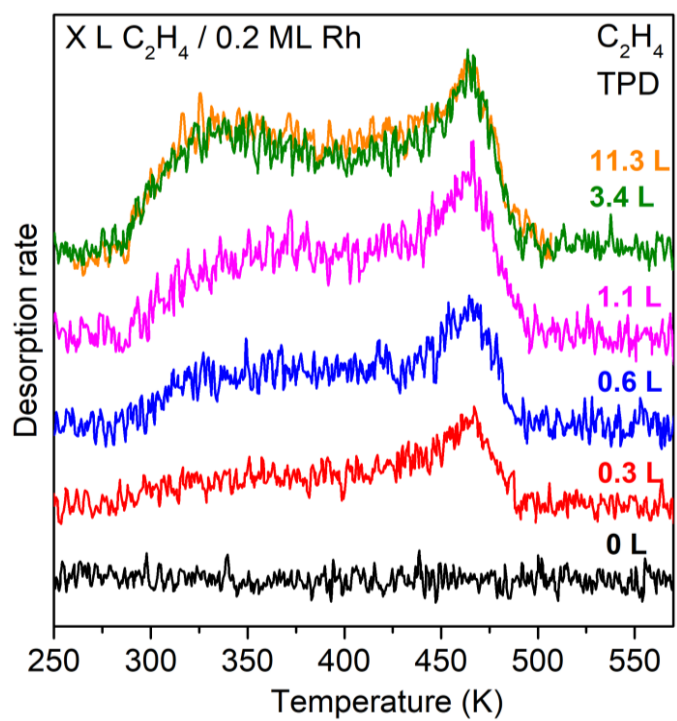

**Figure S3.** A series of $C_2H_4$ TPD results obtained after various $C_2H_4$ exposures at room temperature on a 0.2 ML Rh/$Fe_3O_4$(001) sample (x = 0, 0.3, 0.6, 1.1, 3.4, 11.3 Langmuir). As the dosage of $C_2H_4$ increases, the $C_2H_4$ desorption peaks saturate at 3.4 L $C_2H_4$. This is demonstrated by overlapping the spectra for 3.4 L and a larger dose of 11.3 L $C_2H_4$ at the top of the figure.

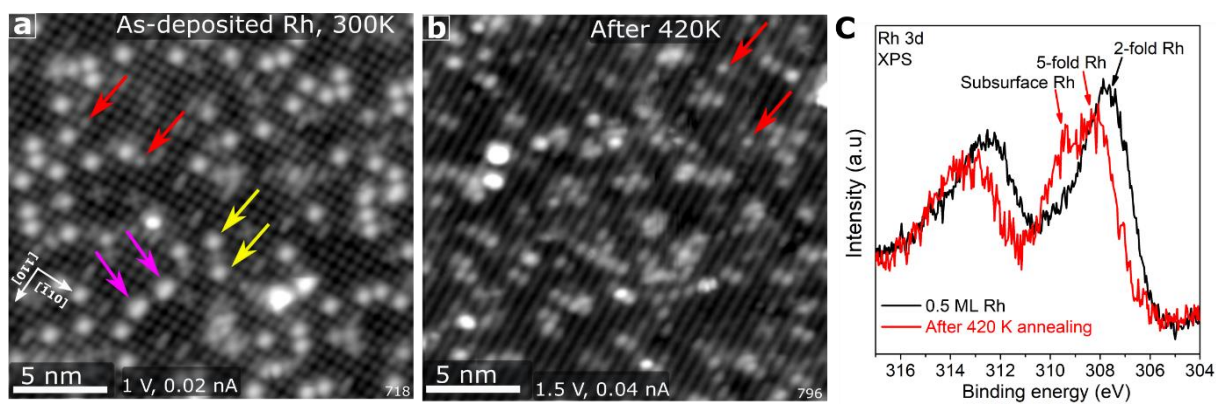

**Figure S4.** Thermal stability of the as-prepared model single-atom $Rh_1/Fe_3O_4(001)$ catalyst. STM images of (a) 0.2 ML Rh on $Fe_3O_4(001)$ after Rh deposition at 300 K and (b) after annealing at 420 K. The yellow arrows indicate 2-fold coordinated $Rh_1$ atoms, which are located in-between the surface Fe rows. The red arrows indicate 5-fold coordinated $Rh_1$ atoms located in the Fe rows along [110] direction. The pink arrows indicate Rh dimer species. (c) XPS Rh 3d collected on the as-prepared Rh sample (black curve) and after annealing at 420 K (red curve). The XPS binding positions of 2-fold Rh, 5-fold Rh, and 6-fold Rh (Rh in subsurface) are pointed out by the black and red arrows.

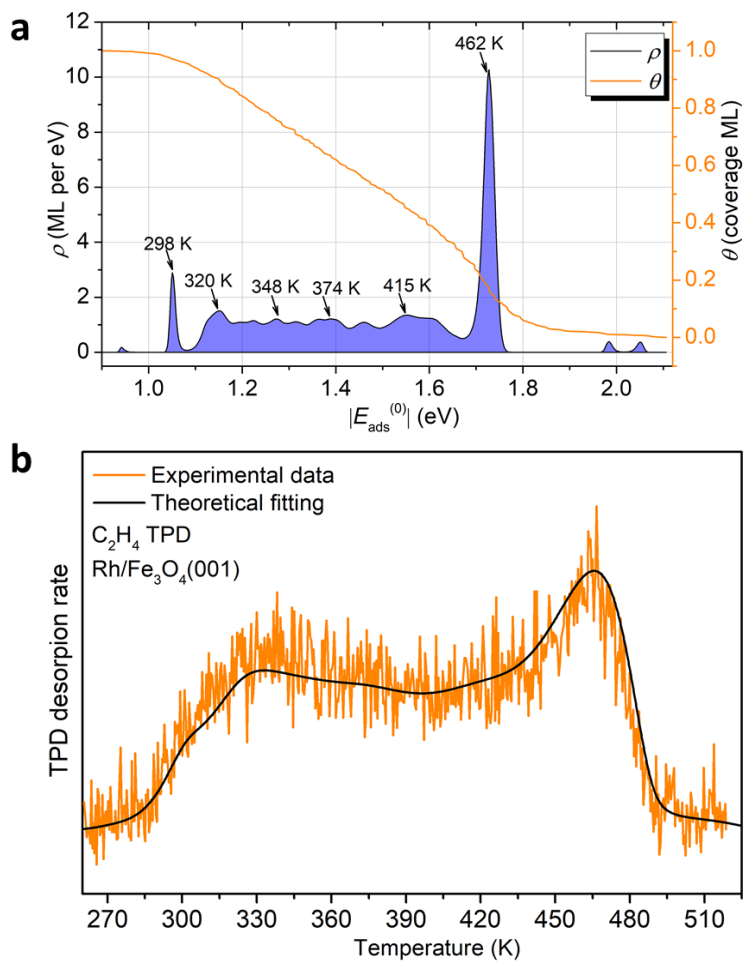

**Figure S5.** Analysis of the TPD spectra using the TPD program introduced recently.[2] (a) Distribution of the adsorption energies derived from the experimental $C_2H_4$ TPD spectra (orange curve in b). The orange curve in (a) shows the fraction of molecules with stronger adsorption than the given value on the x-axis, e.g., the onset of the peak at 462 K is at a coverage between 0.25 and 0.3 of the initial (saturation) coverage. The temperatures in (a) provide a rough indication of the correspondence of the peaks and wiggles in the adsorption energy distribution with features in the spectrum. (b) TPD spectra calculated from the adsorption energy distribution (black curve) in panel (a) plotted on top of the experimental data (orange). The analysis assumes is based on the parameters in Table S1 and results in an adsorption energy of 1.72 eV for the peak at 462 K. A conventional Polanyi−Wigner TPD analysis would yield this result when assuming a pre-exponential factor of $2 \times 10^{18}$ s$^{-1}$.

**Table S1: Input parameters for the TPD analysis**

| $C_2H_4$ gas, atomic coordinates (Å) | C | 0.6579 | −0.0045 | 0.0639 |
|---|---|---|---|---|
| | H | 1.1610 | 0.0661 | 1.0238 |
| | H | 1.3352 | −0.0830 | −0.7815 |
| | C | −0.6579 | 0.0045 | −0.0639 |

|   |   |   |
|---|---|---|
|   | H | −1.3355  0.0830  0.7812 |
|   | H | −1.1608  −0.0661  −1.0239 |
| Heating ramp $\beta$ | 1 K/s | |
| Areal density of adsorption sites $n_a$ | $2.81 \times 10^{17}$ per m$^2$ (0.2 ML Rh) | |
| Initial coverage $\theta_0$ | 1 | |
| Vibration frequencies [a] | C$_2$H$_4$ | 9.57, 17.72, 20.37, 49.50, 60.88, 85.09 meV |
|   | Rh | 18.78, 30.30, 38.47, −10.04, −12.85, −37.71 meV |
| Extra entropy [b] | 2 $k_B$ | |
| Langmuirian initial sticking $s_0$ | 0.06 [c] | |

[a] Frequencies from DFT. Vibrations only present on the bare surface (without the adsorbate) must be entered as negative numbers in the program. The vibrations influence the calculated adsorption energy through the vibrational entropy; omitting the vibrations would lead to an overestimation of the magnitude of the adsorption energy by 0.12 eV.

[b] Two equivalent configurations due to C$_2$H$_4$ flipping on Rh$_1$/Fe$_3$O$_4$(001).

[c] This value corresponds to a sticking cross section of 20 Å² per unoccupied Rh atom. This parameter, which describes the sticking at the desorption temperature, leads to the largest uncertainty of the TPD analysis of ±0.1 eV (assuming an uncertainty by ± one order of magnitude). The appearance of the coverage-dependent TPD curve in Fig. S3 indicates that long-distance adsorbate diffusion on the bare Fe$_3$O$_4$ surface cannot occur; otherwise, the low-coverage spectra should show desorption only at the high-temperature peak. Long-distance diffusion over the Fe$_3$O$_4$ surface would lead to a high sticking coefficient, because the molecules could arrive anywhere on the surface, diffuse and spill over onto the Rh$_1$ adatoms.

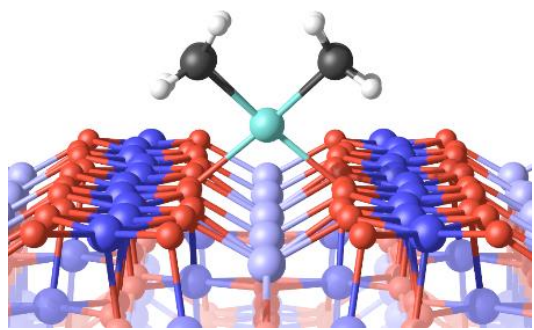

Figure S6. 2C$_2$H$_4$ adsorption on a 2-fold oxygen coordinated Rh$_1$ calculated by DFT. The average adsorption energy per C$_2$H$_4$ molecule is −1.64 eV.